\documentclass[%
 reprint,
 amsmath,amssymb,
 aps,
]{revtex4-2}
\usepackage{braket}
\usepackage{graphicx}
\usepackage{dcolumn}
\usepackage{bm}

\begin{document}

\preprint{APS/123-QED}

\title{Improved Calibration of RF Cavities for Relativistic Electron Beams: Effects of Secondary Corrections and Experimental Verification}
\author{K.~Shih}
\email{kshih@bnl.gov}
\author{I.~Petrushina}
\author{V. N.~Litvinenko}
 \altaffiliation[Also at ]{Brookhaven National Laboratory, Upton, New York 11973, USA}
 \affiliation{Physics Department, Stony Brook University, SUNY, New York 11794, USA}
\author{I.~Pinayev}
\author{J.~Ma}
\author{G.~Wang}
\author{Y.~Jing}
\author{Y.~Wu}
\affiliation{Brookhaven National Laboratory, Upton, New York 11973, USA}

\date{\today}

\begin{abstract}
In the aspect of longitudinal beam bunching, the bunching strength can be controlled by the RF cavity phase and voltage. However, these machine parameters are different from those that interact with the beam itself. In order to gain control of the beam-cavity interaction, cavity calibration must be performed. Furthermore, it relies on fitting the beam energy gain versus cavity phase to a calibration function. Under the conventional assumption of relativistic beam conditions, the calibration function is a first harmonic sinusoidal function (a sinusoidal function with a period of 2$\pi$). However, this expression is insufficient for a high-voltage bunching cavity. Due to beam acceleration inside the cavity, an energy bias and a second harmonic function should be included to modify the conventional calibration function, even for a relativistic electron beam. In this paper, we will derive this modification and provide a comparison to both the Coherent Electron Cooling Experiment and the  IMPACT-T simulation, respectively.

\end{abstract}

\maketitle


\section{\label{sec:level1}Introduction}
A single-cell RF cavity often serves as a bunching cavity for low-energy beamline. In normal design, the bunching cavity frequency, denoted by $f$, is matched with the incoming beam velocity, such that $f = \beta_{0}c/L$, where $L$ is the cavity length. This matching condition allows the incoming beam to ``surf'' on a chosen phase angle of the cavity field. By adjusting the cavity voltage, a designated energy chirp can be introduced to the incoming beam. This energy chirp creates a head-to-tail velocity difference in the beam, causing longitudinal bunching (velocity bunching). Consequently, the target beam current profile can be manipulated by the input phase angle and voltage of the bunching cavity.

However, due to the nature of the RF system, the machine input phase parameters $\theta$ are different from the phase angle of the electric field that the beam is surfing on. It requires calibration to determine cavity parameters, such as the Transit Time Factor (TTF) and the RF phase reference. During calibration, we measure a calibration function, which is defined as the beam energy gain $\Delta U$ after the cavity at different input cavity phases $\theta$, shown in Eq.~(\ref{eq:dU}). Here, $\vec{E}$ denotes the cavity electric field and $\vec{\upsilon}$ denotes the beam velocity.

For relativistic electrons, the effect of velocity change inside the RF cavity is conventionally ignored, leading to the widely accepted first-order calibration function $C^{(1)}(\theta)$, given by Eq.~(\ref{eq:ctheta}). Here, $V_{0}$ denotes the cavity voltage, while $\alpha$ and $\mu_{0}$ denote the TTF and the RF phase reference, respectively.

\begin{figure}[hb!]
\includegraphics[scale=0.42]{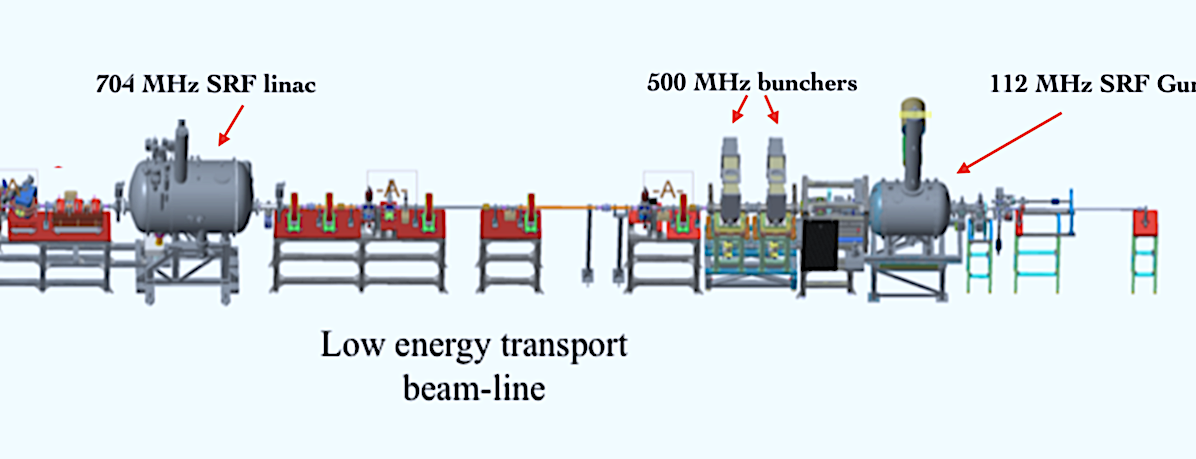}
\caption{\label{fig:cecle}The schematic of the CeC low-energy transportation beamline. The main goal of this section is to prepare a high-quality electron beam for coherent cooling. Currently, only one 500 MHz bunching cavity is in operation.}
\end{figure}

\begin{equation}\label{eq:dU}
\begin{split}
C\left (  \theta \right ) & =  \Delta U \left ( t = L/\beta_{0}c, \theta \right )  \\
&= q \int_{0}^{L/\beta_{0}c} \vec{E} \left (  z \left ( {t}' , \theta \right ) , \theta, {t}' \right )\cdot \vec{\upsilon} \left ( {t}' , \theta \right )d{t}' 
\end{split}
\end{equation}

\begin{equation}\label{eq:ctheta}
C^{(1)}(\theta) = qV_{0}\alpha\sin(\theta-\mu_{0})
\end{equation}

However, Eq.~(\ref{eq:ctheta}) is insufficient for high-voltage RF cavities, such as the single-cell 500MHz bunching cavity used in the Coherent Electron Cooling (CeC) Proof of Principle (PoP) Experiment \cite{Litvinenko_CeC,Pinayev_CeC_Status}. Figure~\ref{fig:cecle} shows the CeC low-energy beam transport (LEBT) section, where the electron bunches ($\sim$1.5 nC, $\sim$380 ps) are generated in the 113MHz quarter-wave superconducting RF (SRF) photoinjector \cite{petrushina2019measurements, petrushina2020high}, which provides a primary acceleration of 1.25MeV. Before the beam reaches the 704 MHz SRF 5-cell linac for the final energy boost of 13.1 MeV, the desired beam peak current is achieved by means of longitudinal bunching. The required compression is obtained through the use of a normal-conducting 500~MHz bunching cavity \cite{buncherpaper} ($L =$ 0.57 m) that is operated at 185 kV, providing $\sim$20 times the bunch length compression in a drifting course of 10 meters.

During cavity calibration, both through experiment and simulations (using IMPACT-T \cite{qiang2006three}), for the CeC relativistic electron beam, we observed significant deviation from the conventional calibration function shown in Eq.~(\ref{eq:ctheta}). For example, the bunching phase reference can shift from $\sim$ 0.2 to 0.5 degrees for the setting of $\sim$ 100 kV to 200 kV bunching voltage. These deviations strongly affected our machine optimization for beam emittance and peak current control. A higher-order correction in the beam-cavity interaction is needed to achieve controllable beam instability.

While the higher-order correction in the beam-cavity interaction was realized and derived in the late 1980s for low-velocity ions under Newtonian mechanics \cite{delayen1987longitudinal, carne1970numerical},  it is currently unclear whether this effect has been systematically measured and studied for relativistic electron beams. Although it is possible that this problem has been observed and addressed in other electron linacs, to our knowledge, no comprehensive study or model has been developed for correcting this effect beyond the first-order formula in the case of relativistic electron beams. In the following sections, we will derive the second-order correction to the calibration function using an alternative approach. By expanding the phase-space parameters of the traveling beam in terms of energy gain, we arrived at a relativistic version of the modified calibration function with corrections (compared with the result in Refs.~\cite{delayen1987longitudinal}). We will show detailed derivation steps in the appendix sections. Furthermore, we will provide a direct comparison of the theoretical model with experimental measurement, IMPACT-T simulation results, and the non‐relativistic model of Refs.~\cite{delayen1987longitudinal}.

\section{Theoretical Model}
To solve Eq.~(\ref{eq:dU}), we expand both the beam velocity $\upsilon\left ( t, \theta \right )$ and the distance traveled by the beam $z\left ( t, \theta \right )$ with respect to the change in beam energy caused by the bunching cavity, $\Delta U\left ( t, \theta \right )$, as shown in Eq.~(\ref{eq:vzexpand}). Furthermore, we use the Fourier form of $E_{z}$ in Eq.~(\ref{eq:Ez}) to represent an arbitrary RF cavity field, where $a_{k}$ and $\phi_{k}$ are the Fourier coefficients. 

\begin{equation}\label{eq:vzexpand}
\begin{split}
\upsilon\left ( t, \theta \right )   & = \beta_{0} c + \frac{\beta_{0}^{-1} \gamma_{0}^{-3}}{mc}  \Delta U\left ( t, \theta  \right )  + \cdots \\
z\left ( t, \theta  \right )  & = \beta_{0} ct + \frac{\beta_{0}^{-1} \gamma_{0}^{-3}}{mc} \int_{0}^{t}  \Delta U \left ( {t}', \theta  \right ) d{t}' + \cdots
\end{split}
\end{equation}

\small
\begin{equation}\label{eq:Ez}
E_{z}\left (  z, t, \theta \right ) = E_{0} \sum_{k = 0}^{N} a_{k} \cos\left ( \frac{k\pi}{L} z - \phi_{k}   \right )\cos\left ( 2\pi f t  + \theta  \right )
\end{equation}
\normalsize

\subsection{First Order Calibration Function}

Under the assumption of low bunching voltage for a relativistic incoming beam, only the first order of $\upsilon$ and $z$ is kept for the buncher calculation, shown in Eq.~(\ref{eq:1thvz}). As a result, at any given time $t$, the beam energy gain can be written as Eq.~(\ref{eq:dU1}), where $\varepsilon_{\upsilon}$ and $\tau_{\upsilon}$ are functions of time $t$. If we set the time equal to $L/\beta_{0}c$, we obtain the conventional calibration function shown in Eq.~(\ref{eq:ctheta}).

\begin{equation}\label{eq:1thvz}
\begin{split}
\upsilon^{\left ( 1  \right ) }\left ( t, \theta \right ) & =  \beta_{0} c   \\
z^{\left ( 1  \right ) }\left ( t, \theta  \right ) & =  \beta_{0} ct 
\end{split}
\end{equation}

\begin{equation}\label{eq:dU1}
\begin{split}
\Delta U^{(1)} \left ( t, \theta  \right ) & =   \gamma_{0}^{3}m\beta_{0}c\varepsilon_{\upsilon} \left ( t  \right )\cos\left [ \theta -  \tau_{\upsilon}\left ( t  \right ) \right ]
\end{split}
\end{equation}

\begin{figure}
\includegraphics[scale=0.25]{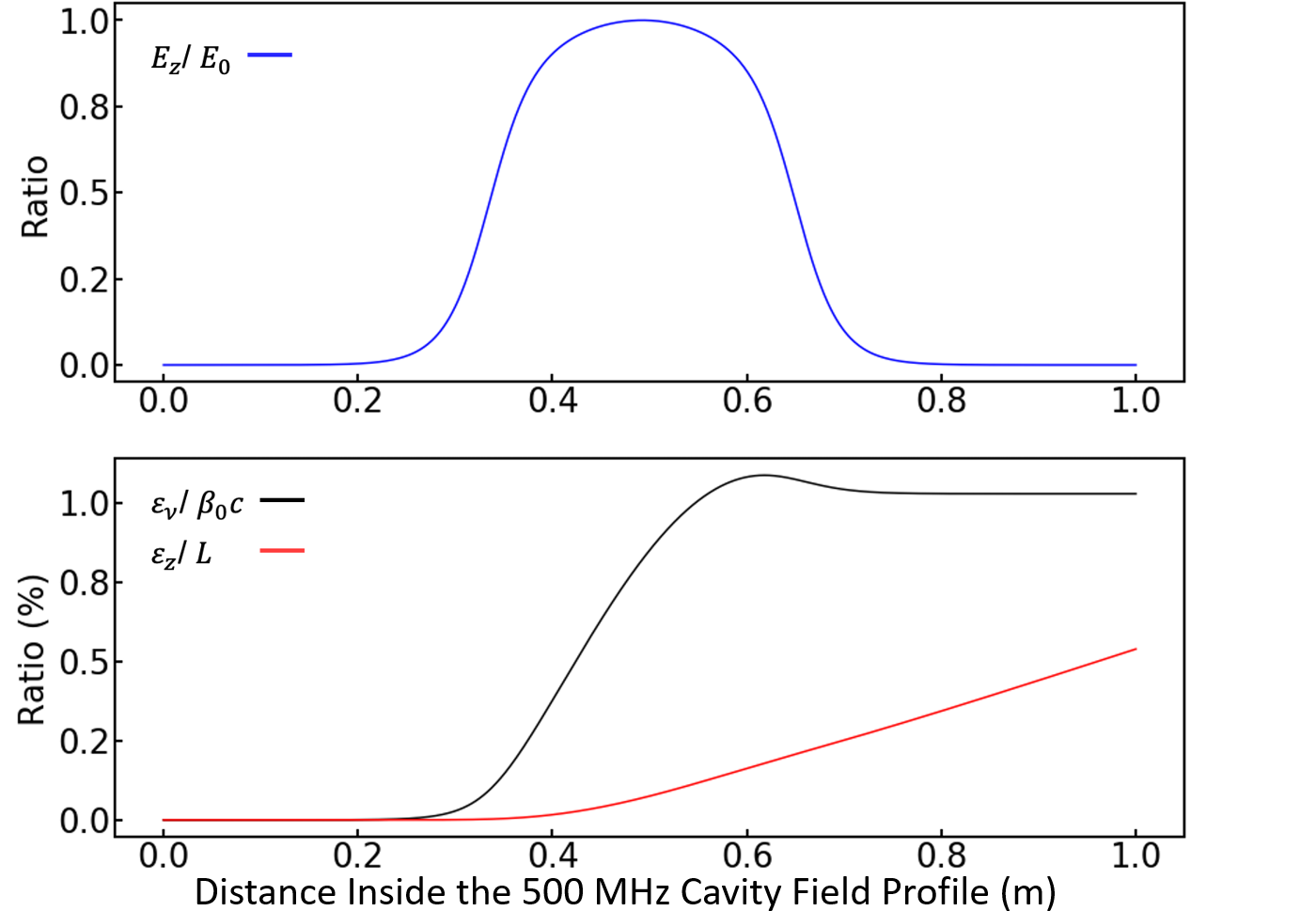}
\caption{\label{fig:evez}The upper plot shows the central electric field of the CeC 500 MHz bunching cavity, while the lower plot displays the evolution of the parameters $\frac{\varepsilon_{\upsilon}}{\beta_{0}c}$ (red line) and $\frac{\varepsilon_{z}}{L}$ (black line) inside the bunching cavity. Both $\varepsilon_{\upsilon}$ and $\varepsilon_{z}$ are approximately 1\% of their leading order terms.}
\end{figure}

\subsection{Second Order Calibration Function}

On the other hand, at high bunching voltage, the calibration function should include beam acceleration. By substituting Eq.~(\ref{eq:dU1}) into Eq.~(\ref{eq:vzexpand}), we obtain the second-order correction of beam velocity and beam distance traveled for an accelerating beam characterized by two time functions $\varepsilon_{\upsilon}(t)$ and $\varepsilon_{z}(t)$, shown in Eq.~(\ref{eq:2thvz}).

\begin{equation}\label{eq:2thvz}
\begin{split}
\upsilon^{(2)}\left ( t, \theta \right )   & = \varepsilon_{\upsilon} \left ( t  \right )\cos\left ( \theta -  \tau_{\upsilon}\left ( t  \right ) \right ) \\
z^{(2)}\left ( t, \theta  \right )  & = \varepsilon_{z} \left ( t  \right )\cos\left ( \theta -  \tau_{z}\left ( t  \right ) \right )
\end{split}
\end{equation}

\begin{equation}\label{eq:dU2}
 \Delta U^{(2)}\left ( t, \theta \right )  =\delta_{0}\left ( t  \right )  +\delta_{1} \left ( t  \right ) \cos\left ( 2\theta -  \tau_{v2}\left ( t  \right ) \right )
\end{equation}
 
In general, $\varepsilon_{z}$ and $\varepsilon_{\upsilon}$ are smaller than their leading-order parameters $L$ and $\beta_{0}c$ at any given location inside the bunching cavity, as shown in Fig.\ref{fig:evez}. Therefore, we expand the integral in Eq.~(\ref{eq:dU}) up to the first order of $\varepsilon_{z}$ and $\varepsilon_{\upsilon}$ without losing generality. After the approximation, we obtain an expression for the second-order correction in beam energy gain $\Delta U^{(2)}$, shown in Eq.~(\ref{eq:dU2}). Where the newly appeared time functions $\delta_{0}\left ( t \right )$ and $\delta_{1} \left ( t \right )$ in this equation are also independent of the phase angle $\theta$. Unlike the form of $\Delta U^{(1)}$, $\Delta U^{(2)}$ has no first harmonic function term, e.g., $\cos(\theta-\mu)$, but has a bias term and a second harmonic function. Both terms are suppressed by the factor $\frac{q^{2}V^{2}
_{0}\alpha^{2}}{\gamma_{0}^{2}P_{0}c}$ at the end of the bunching cavity (when $t = L/\beta_{0}c$), where $P_{0}$ is the initial beam momentum. As a result, the second-order correction of the calibration function has a quadratic dependence on the bunching voltage. This result is consistent with the second-order longitudinal transit time factor derived in Refs.~\cite{delayen1987longitudinal} for slow-moving ions ($\beta$ below 0.1), which did not take the relativistic effect into account.

\small
\begin{equation}\label{eq:C01}
\begin{split}
C\left (\theta  \right ) & =qV_{0}\alpha \sin\left ( \theta -  \mu_{0} \right ) + \frac{q^{2}V^{2}_{0}\alpha^{2}}{\gamma_{0}^{2}P_{0}c} \left [  \sigma_{0}  +  \sigma_{1} \sin\left ( 2\theta -  \mu_{1} \right ) \right ] \\
& where \: \delta_{i}\left (L / \beta_{0}c   \right ) = \frac{q^{2}V^{2}_{0}\alpha^{2}}{\gamma_{0}^{2}P_{0}c} \sigma_{i} ; \; \; \;i = 0, 1
\end{split}
\end{equation}
\normalsize

\begin{equation}\label{eq:zegp}
\theta_{0} \approx \mu_{0} - \frac{qV_{0}\alpha}{\gamma_{0}^{2}P_{0}c} \left [  \sigma_{0}  +  \sigma_{1} \sin\left ( 2\mu_{0} -  \mu_{1} \right ) \right ] 
\end{equation}

Lastly, Eq.~(\ref{eq:C01}) shows the modified calibration function, where $\sigma_{0}$ and $\sigma_{1}$ are dimensionless parameters related to the cavity field structure. In a setup with relatively high bunching voltage, the bias term in Eq.~(\ref{eq:C01}) can cause a significant shift in beam energy. If the conventional calibration function is used, this energy shift may be incorrectly interpreted as an increase in the initial beam energy. Additionally, the reference phase $\theta_{0}$ (phase of zero beam energy gain) is also shifted from the conventional RF
phase reference $\mu_{0}$ due to the effect of beam acceleration. This phase shift has a rough linear relation with the bunching voltage, as shown in Eq.~(\ref{eq:zegp}). Consequently, using only Eq.~(\ref{eq:ctheta}) to calibrate a high voltage bunching cavity can compromise the accuracy of beam dynamics calculations.


\begin{figure}
\includegraphics[scale=0.6]{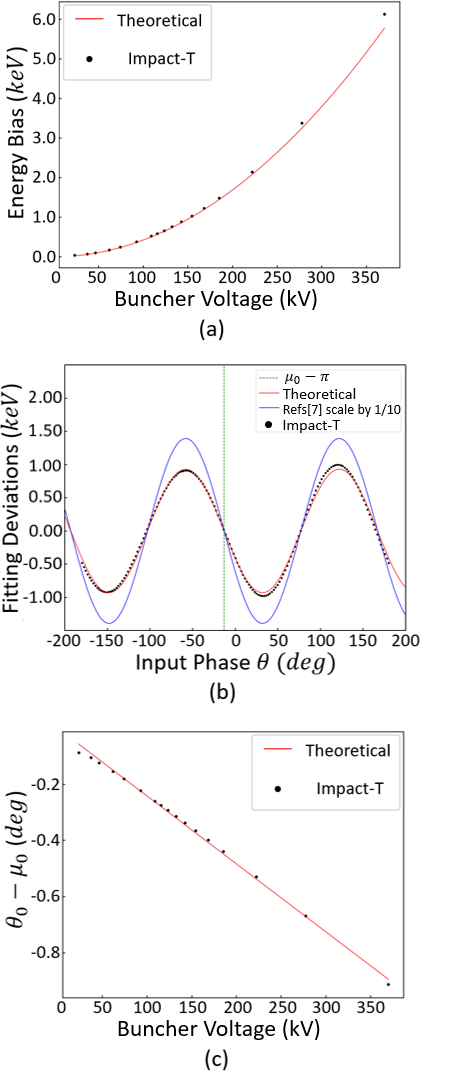}
\caption{\label{fig:constantshift}The comparison between  IMPACT-T simulation (black dots) and theoretical model (red line) of the CeC 1.25 MeV electron beam is shown in the following three plots: (a) compares the bias term in Eq.~(\ref{eq:C01}); (b) compares the second harmonic term in Eq.~(\ref{eq:C01}) and result calculated by means of Refs.~\cite{delayen1987longitudinal} (blue line); and (c) compares the zero energy gain phase $\theta_{0}$, which is often used as a reference phase during bunching cavity calibration.}
\end{figure} 

\section{Simulation Comparison}

To verify our modification, we compared Eq.~(\ref{eq:C01}) and Eq.~(\ref{eq:zegp}) with a CeC  IMPACT-T simulation model \cite{jing2021beam}. During the simulation, we propagated the CeC electron beam through a 500 MHz bunching cavity with variant voltage, ranging from 8 kV to 360 kV (initial beam energy is 1.25 MeV), and recorded the exiting beam energy. At each bunching cavity voltage, we scanned a complete $2\pi$ cavity phase in 160 steps to obtain a single calibration function. These simulated calibration functions were then individually fitted with a general function, e.g., $ C\left (\theta \right ) = A+B\sin(\theta+a)+C\sin(2\theta + b)$. We plotted all the fitting constants $A$ against their bunching voltage in Fig.\ref{fig:constantshift}(a), together with the theoretical expression of the bias term shown in Eq.~(\ref{eq:C01}) for comparison. 


Figure~\ref{fig:constantshift}(b) shows a comparison of the second-order harmonic term obtained from our modified calibration function and the formula from Refs.~\cite{delayen1987longitudinal}, which without considering the relativistic effects. Notably, our modified calibration function with the relativistic correction term shows much better agreement with the  IMPACT-T simulated data. This clearly demonstrates the significance of our proposed modification. We also compared the phase shift obtained from our modified calibration function with the simulated results, as shown in Fig.\ref{fig:constantshift}(c). Overall, the  IMPACT-T simulation of the CeC experiment matched well with our modified calibration function.

However, one may notice some discrepancies in both Fig.\ref{fig:constantshift}(a) and Fig.\ref{fig:constantshift}(b). These discrepancies are the result of the contribution of higher-order corrections (e.g., the third-order correction), which are outside the scope of this paper and the current CeC experiment.

\section{Experimental Comparison}
\begin{figure}
\includegraphics[scale=0.18]{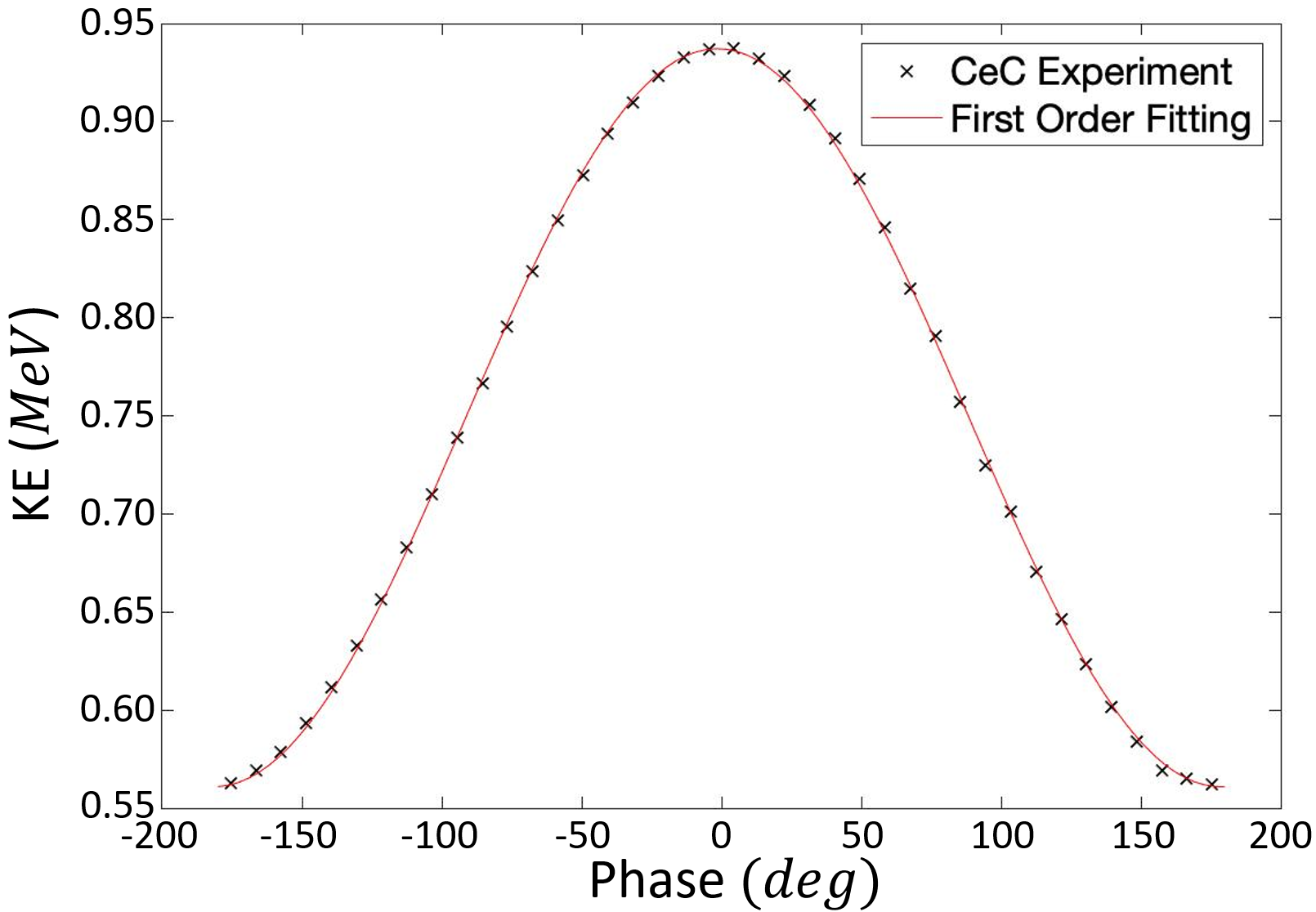}
\caption{\label{fig:CeCraw}This data represents a complete CeC 500 MHz bunching cavity calibration, in which the beam energy $U(\theta)$ was measured downstream with different phase angle settings $\theta$. The measurement data is shown as black crosses, and a first-order fitting using Eq.~(\ref{eq:ctheta}) is displayed as the red line.}
\end{figure}

\begin{figure}
\includegraphics[scale=0.26]{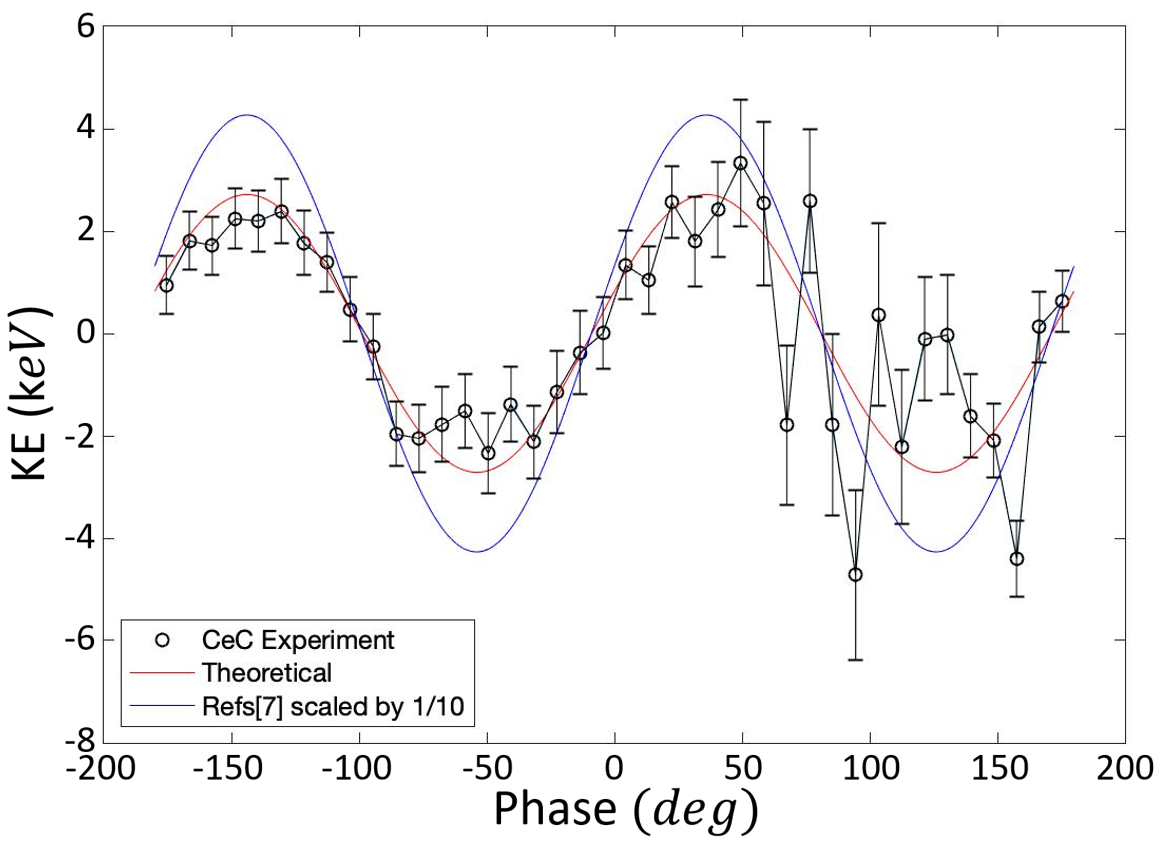}
\caption{\label{fig:TCeCd2}The comparison between the theoretical model of the second-order calibration function and the experimental measurement. The experimental data from the CeC experiment is represented as black circles, while our theoretical model is displayed as a red line and Refs.~\cite{delayen1987longitudinal} formula as a 
blue line.}
\end{figure}

Moreover, we also observed similar deviations in the calibration function during the CeC experiment. Figure~\ref{fig:CeCraw} shows the data from a complete calibration of the CeC 500 MHz bunching cavity. Each point of $U(\theta)$ on Fig.~\ref{fig:CeCraw} was obtained by a single CeC beam energy measurement individually, using the downstream trims, solenoid, and YAG screen \cite{brutus2014coherent, igorEmeasurement}. In this particular calibration, the initial CeC electron beam energy was 0.75 MeV, and the bunching cavity was set to around 190 kV. After subtracting the bias and first harmonic component from the measurement data, we obtained the higher-order components (dominated by the second-order harmonic) of the calibration function. The comparison of the subtracted result, the theoretical prediction from Eq.(\ref{eq:C01}), and the formula of Refs.~\cite{delayen1987longitudinal} is shown in Fig.\ref{fig:TCeCd2}. The measurement data exhibited a good match in terms of amplitude and frequency with our theoretical prediction. However, the formula of Refs.~\cite{delayen1987longitudinal} showed a significant discrepancy, which is mainly attributed to the lack of relativistic gamma factor suppression in Newtonian mechanics. This causes an overestimation of the effect in the case of a relativistic electron beam, such as the one in the CeC accelerator.


However, a phase-dependent discrepancy between the experimental measurement and our theoretical model can also be identified from Fig.\ref{fig:TCeCd2}. This small discrepancy can be caused by even higher-order corrections (such as the third-order harmonic) of the calibration function, convoluted with machine errors. The most common machine errors are beam energy fluctuations and time jitter of the RF system (for the CeC beam, $\sim \pm$ 10 ps jitter and $\sim \pm$ 0.1 keV energy fluctuations). Isolating the effect of higher-order correction is challenging since each single energy measurement requires at least 32 CeC electron beam bunches to reach the desired measurement accuracy. Fig.\ref{fig:TCeCd2} contains a total of 1280 measurements of individual CeC electron beams taken over a period of four hours. To measure only the effect of higher-order calibration functions requires extremely high stability from the CeC machine.

\section{Conclusion}

In conclusion, the acceleration inside the bunching cavity induces a secondary correction to the conventional calibration function, which is significant even for relativistic electron beams. This correction creates an additional energy bias and adds a second harmonic function to the conventional calibration function. Both additional terms have a quadratic dependence on the bunching voltage. Additionally, these corrections induce a linear phase shift to the reference phase of the conventional calibration function, which can affect downstream beam dynamics, including emittance and peak current.

Notably, the formula of Refs.~\cite{delayen1987longitudinal}, developed for low-velocity beams, overestimates the corrections for our intermediate energy beams. This finding emphasizes the significance of our proposed modification, which takes into account the relativistic effects and provides a more accurate calibration function for high-voltage bunching systems. In the example of the CeC 500 MHz bunching cavity, which operates at a voltage of 185 kV, the electron beam with an initial energy of 1.25 MeV will be shifted by a 1.6 keV energy bias after the bunching cavity. Furthermore, the reference phase of the cavity will also be shifted by 0.46 degrees. These modifications improve the accuracy of the downstream beam dynamics control and are crucial for optimizing the performance of relativistic electron beams in various applications.

\begin{acknowledgments}
We would like to thank the RHIC operations team and the RHIC C-AD instrumentation group, in particular P. Inacker, for their support concerning the RF system and injection laser system. This work was supported by the Coherent Electron Cooling (CeC)
Proof of Principle Experiment.
\end{acknowledgments}

\appendix

\section*{Appendix}
It is well known that the first-order cavity calibration function is a first harmonic function of the cavity phase $\theta$, causing the second-order beam velocity to also be a first harmonic function. Therefore, it is convenient to express our physical quantities on the odd-even Fourier basis to derive the higher-order beam energy gain (energy operators in $t$ and $\theta$ are separated). The electric field can be expressed as a rotation operator in time and upping and lowering operator in $\theta$. Equation (\ref{eq:v_f_basis}) and Eq.~(\ref{eq:DU_basis}) show the expression of the beam velocity and beam energy gain. 

\small
\begin{equation}\label{eq:v_f_basis}
\begin{split}
    &~~~~\upsilon\left ( t, \theta \right ) \\
    & = \sum_{m }\bra{\pm}\left[\upsilon_{m}\left(t\right)\right]\ket{\pm,m}\\
    & = \sum_{m}\begin{bmatrix}1&1\end{bmatrix}\left[\upsilon_{m}\left(t\right)\right]\begin{bmatrix}\sin\left(m \theta\right)\\\cos\left(m \theta\right)\end{bmatrix}
\end{split}
\end{equation}

\small
\begin{equation}\label{eq:DU_basis}
\begin{split}
    &~~~~\Delta U_{m}\left( t, \theta \right )\\  
    & = \bra{\pm}\int^{t}_{0}dt'\left[E_{z}\left (  z(t'), t', \theta \right )\right] \left[\upsilon_{m}\left(t'\right)\right] \ket{\pm,m}\\
    & = \sum_{k}\left.\int^{t}_{0}dt' \frac{qE_{0}a_{\left|k  \right|}}{2}\bra{\pm}\right\{\\
    &~~~~~~~\cos\left(\psi_{k}\left(t'\right)\right)\cos\left( \theta-\tilde{k}\phi_{\left|k  \right|}\right)\left[\upsilon_{m}\left(t'\right)\right]\ket{\pm,m}\\
    &~~~~+\left.\sin\left(\psi_{k}\left(t'\right)\right)\sin\left( \theta-\tilde{k}\phi_{\left|k  \right|}\right)\left[\upsilon_{m}\left(t'\right)\right]\ket{\pm,m}\right\}\\    
    & =\sum_{h}^{-1,~+1} \bra{\pm,1+hm}\left[\Delta U_{m,h}\left(t\right)\right]^{T}\ket{\pm}\\
\end{split}
\end{equation}
\small
\begin{equation}\label{eq:U_T}
\begin{split}
    ~\mbox{where:~~~}&\\
    &~~~\left[\Delta U_{m,h}\left(t\right)\right]^{T}\ket{\pm}\\ &=\sum_{k}\frac{qE_{0}a_{\left|k\right|}}{4}\left[\mathrm{R}\left(-\tilde{k}\phi_{\left|k\right|}\right)\right]\left.\int^{t}_{0}dt'\right\{\\
    &~~\left.\left[\mathrm{R}\left(\psi_{k}\left(z(t'),t'\right)\right)\right]\left[-\sigma_{z}\right]^{\frac{1-h}{2}}\left[\upsilon_{m}\left (t'\right ) \right]\ket{\pm}\right\} \\
\end{split}
\end{equation} 

\small
\begin{equation}\label{eq:R_w}
\begin{split}       
    \left[\mathrm{R}\left(\psi_{k}\right)\right] &= \begin{bmatrix}\cos\left(\psi_{k}\right)&-\sin\left(\psi_{k}\right)  \\\sin\left(\psi_{k}\right)&\cos\left(\psi_{k}\right) \end{bmatrix}\\
    \psi_{k}\left(z(t),t\right) &= 2\pi f t+k \frac{\pi}{L}z(t)
\end{split}
\end{equation}

The Pauli matrices are denoted by $\sigma$ (i.e., $\sigma_{z}$ represents $\sigma$ in the $z$ direction). The summation on $k$ has been extended to $k=-N,\cdots, 0,0,\cdots,N$; where $\tilde{k}$ is the sign function of $k$, and $h$ only takes two values, -1 and 1. Equation (\ref{eq:U_T}) can be easily proven by substituting Eq.~(\ref{eq:v_f_basis}) into Eq.~(\ref{eq:dU}). Furthermore, Equations (\ref{eq:U_T}) and (\ref{eq:vzexpand}) provide a framework for calculating beam energy gain perturbation under the basis of $\theta$ harmonics, which also includes relativistic effects. Additionally, for any beam velocity vector $\ket{\pm,m}$, it will always create a lower beam energy vector $\ket{\pm,n-m}$ and an upper beam energy vector $\ket{\pm,n+m}$, where $n$ is the electric field harmonic number in $\theta$, and in our discussion, $n=1$. Since the first-order beam energy gain is $\ket{\pm,1}$, the second-order energy gain must contain a bias energy $\ket{\pm,0}$ and a second-harmonic energy $\ket{\pm,2}$. The calculation of the first-order beam energy gain is presented below.

\small
\begin{equation}\label{eq:U1_basis}
\begin{split}
    &~~~\Delta U^{(1)}\left(t\right) \\
    & =\sum_{h} \bra{\pm, 1+0}\left[\Delta U_{0,h}\left(t\right)\right]^{T}\ket{\pm} \\
    & = \bra{\pm, 1}\sum_{k}\frac{qE_{0}}{4}a_{\left|k\right|}\left[\mathrm{R}\left(-\tilde{k}\phi_{\left|k\right|}\right)\right]\left.\int^{t}_{0}dt'\right\{\\
    &~~~~~~~~~\left.\left[\mathrm{R}\left(2\pi f t'+k \frac{\pi}{L}z^{(1)}\right)\right]\left[I-\sigma_{z}\right]\left[\upsilon^{(1)}_{0}\right]\ket{\pm}\right\} \\
    & = \bra{\pm, 1}\sum_{k}\frac{qE_{0}\beta_{0}c}{2}a_{\left|k\right|}\left[\mathrm{R}\left(-\tilde{k}\phi_{\left|k\right|}\right)\right]\int^{t}_{0}dt'    \left[\mathrm{R}\left(\omega_{k}t'\right)\right] \begin{bmatrix}0 \\ 1\end{bmatrix} \\    
    & =  \gamma_{0}^{3}m\beta_{0}c \bra{\pm, 1}  \begin{bmatrix}\varepsilon_{\upsilon s}\left(t\right) \\ \varepsilon_{vc}\left(t\right)\end{bmatrix} 
\end{split}
\end{equation}
\small
\begin{equation}\label{eq:evsc_basis}
\begin{split}
    \mbox{where:~} ~~~& \\
    \omega_{k} & = 2\pi f +k \frac{\pi}{L}\beta_{0}c \\
    \begin{bmatrix}\varepsilon_{\upsilon s}\left(t\right) \\ \varepsilon_{vc}\left(t\right)\end{bmatrix} & = \sum_{k}\frac{qE_{0}}{2\gamma_{0}^{3}m\omega_{k}} a_{\left|k\right|}\left[\mathrm{R}\left(-\tilde{k}\phi_{\left|k\right|}\right)\right]\begin{bmatrix}\cos\left(\omega_{k}t\right)-1 \\ \sin\left(\omega_{k}t\right)\end{bmatrix}
\end{split}
\end{equation}

Using Eq.~(\ref{eq:vzexpand}), we can obtain the second-order beam velocity and distance traveled.
\small
\begin{equation}\label{eq:mix}
\begin{split}
    \upsilon^{(2)}\left(t\right) &= \bra{\pm, 1}  \begin{bmatrix}\varepsilon_{\upsilon s}\left(t\right) \\ \varepsilon_{vc}\left(t\right)\end{bmatrix}\\
    z^{(2)}\left(t\right) &=\bra{\pm, 1}  \int^{t}_{0}dt'\begin{bmatrix}\varepsilon_{\upsilon s}\left(t'\right) \\ \varepsilon_{vc}\left(t'\right)\end{bmatrix}\\
    &= \bra{\pm, 1}  \begin{bmatrix}\varepsilon_{zs}\left(t\right) \\ \varepsilon_{zc}\left(t\right)\end{bmatrix}
\end{split}
\end{equation}

\begin{equation}\label{eq:ez2sc_basis}
\begin{split}
    \mbox{where:~} ~~~~& \\
    \begin{bmatrix}\varepsilon_{zs}\left(t\right) \\ \varepsilon_{zc}\left(t\right)\end{bmatrix} & = \sum_{k}\frac{qE_{0}}{2\gamma_{0}^{3}m\omega_{k}^{2}} a_{\left|k\right|}\left[\mathrm{R}\left(-\tilde{k}\phi_{\left|k\right|}\right)\right]\begin{bmatrix}\sin\left(\omega_{k}t\right)-\omega_{k}t \\ 1-\cos\left(\omega_{k}t\right)\end{bmatrix}\\
\end{split}
\end{equation}

Consequently, we can express these second-order beam quantities in the polar format, with $\varepsilon$ being the vector length and $\tau$ being the vector angle(e.g., $\varepsilon_{\upsilon}=\sqrt{\varepsilon_{\upsilon s}^{2}+\varepsilon_{\upsilon c}^{2}}$,  $\tan\left[\tau_{\upsilon}\left ( t \right ) \right]= \varepsilon_{\upsilon s}\left(t\right)/\varepsilon_{\upsilon c}\left(t\right)$).

\begin{equation}\label{eq:vz2_o}
\begin{split}
\upsilon^{(2)}\left(t\right) &= \varepsilon_{\upsilon}\cos\left(\theta-\tau_{\upsilon}\left ( t  \right )\right)\\
z^{(2)}\left(t\right) &= \varepsilon_{z}\cos\left(\theta-\tau_{z}\left ( t  \right )\right)\\
\end{split}
\end{equation}

Furthermore, we can define the TTF $\alpha$ and the RF phase reference $\mu_{0}$ as follows. 

\begin{equation}\label{eq:a_mu0}
\begin{split}
    Q &= \varepsilon_{\upsilon} \left ( L/\beta_{0}c  \right )\frac{\gamma_{0}^{3}m}{qE_{0}} \\
    \alpha &= \frac{\beta_{0}cE_{0}}{V_{0}}Q \\
    \mu_{0} &= \tau_{\upsilon}\left ( L/\beta_{0}c  \right ) - \pi/2   
\end{split}
\end{equation} 

By including Eq.~(\ref{eq:mix}) into Eq.~(\ref{eq:U_T}), we can obtain the beam energy gain up to the second order.

\small
\begin{equation}\label{eq:UT_basis}
\begin{split}
    &~~~\Delta U\left(t\right)\\
    & = \sum_{m,h} \bra{\pm, 1+ h m}\left[\Delta U_{m,h}\left(t\right)\right]^{T}\ket{\pm} \\
    & \approx \sum_{m,h,k}\bra{\pm, 1+ h m}\frac{qE_{0}}{4}a_{\left|k\right|}\left[\mathrm{R}\left(-\tilde{k}\phi_{\left|k\right|}\right)\right]\left.\int^{t}_{0}dt'\right\{ \\
    & ~~~\left.\left[\mathrm{R}\left(2\pi f t'+k \frac{\pi}{L}\left(z^{(1)}+z^{(2)}\right)\right)\right]\left[-\sigma_{z}\right]^{\frac{1-h}{2}}\left[\upsilon^{(1)}_{m}+\upsilon^{(2)}_{m}\right]\ket{\pm}\right\} \\
    & \approx \cdots \left[\mathrm{R}\left(\omega_{k} t'\right)\right]\left(\left[I\right]+ k\pi\frac{ z^{(2)}}{L}\left[-i\sigma_y\right]\right)\left[-\sigma_{z}\right]^{\frac{1-h}{2}}\left[\upsilon^{(1)}_{m}+\upsilon^{(2)}_{m}\right]\ket{\pm}\\
    &= \Delta U^{(1)}\left(t\right) + \sum_{m,h,k}\bra{\pm, 1+ hm}\cdots \left\{\left[\mathrm{R}\left(\omega_{k} t'\right)\right]\left[-\sigma_{z}\right]^{\frac{1-h}{2}}\left[\upsilon^{(2)}_{m}\right]\right.\\
    &~~~~~~~~~~~~~~~~  + \left.k\pi\left[\mathrm{R}\left(\omega_{k} t'\right)\right]\left[\sigma_{z}\right]^{\frac{1-h}{2}}\left[-i\sigma_y\right]\left[\left(\frac{z^{(2)}}{L}\upsilon^{(1)}\right)_{m}\right]\right\}\ket{\pm}\\
    &= \Delta U^{(1)}\left(t\right) + \sum_{h,k}\bra{\pm, 1+ h}\cdots \left[\mathrm{R}\left(\omega_{k} t'\right)\right]\left[-\sigma_{z}\right]^{\frac{1-h}{2}}\begin{bmatrix}\varepsilon_{\upsilon s}\left(t'\right) \\ \varepsilon_{vc}\left(t'\right)\end{bmatrix}\\
    & ~~~~~~~~ + \sum_{h,k}\bra{\pm, 1+ h}\cdots \frac{k\pi\beta_{0}c}{L}\left[\mathrm{R}\left(\omega_{k} t'\right)\right]\left[\sigma_{z}\right]^{\frac{1-h}{2}}\begin{bmatrix}-\varepsilon_{zc}\left(t'\right) \\ \varepsilon_{zs}\left(t'\right)\end{bmatrix}\\    
    &= \Delta U^{(1)}\left(t\right) +\sum_{h}\bra{\pm, 1+ h}\left[\delta_{\frac{1+h}{2}}\left(t\right)\right]
\end{split}
\end{equation}

By relaxing the integration boundary from $\{0,t\}$ to $\{s,t\}$, we can express the second-order beam energy gain in terms of the rotational operator $R$, as shown in Eq.~(\ref{eq:U2_basis}). Similarly to Eq.~(\ref{eq:vz2_o}) and Eq.~(\ref{eq:a_mu0}), we can define the parameters $\delta_{0}$, $\delta_{1}$, and $\mu_{1}$.
\\
\begin{equation}\label{eq:U2_basis}
\begin{split}
    &~~~\left[\Delta U_{h}^{(2)}\left(s,t\right)\right]\\
    &=\left[\delta_{\frac{1+h}{2}}\left(s,t\right)\right]\\
    &= \frac{q^{2}V^{2}_{0}\alpha^{2}}{\gamma_{0}^{3}m \beta_{0} c^{2} }\sum_{k,j} \frac{a^{2}_{\left|k\right|\left|j\right|}}{8\omega_{j}Q^{2}}\left[\mathrm{R}\left(-\varphi_{h,\left|k\right|\left|j\right|}\right)\right]\cdot\left.\int^{t}_{s}\int^{t'}_{s}dt'dt''\right\{\\  
    &~~~~~~~~~(\frac{\omega_{j}}{\beta_{0}}+h \frac{k\pi c}{L})\left[\mathrm{R}\left(\omega_{k} t' +h \omega_{j} t''\right)\right]\\
    &~~~~~~~~~~~~~~~- \left.h\frac{k\pi c}{L}\left[\mathrm{R}(\omega_{k} t'+h\omega_{j}s)\right]\right\}\cdot\begin{bmatrix}0\\1\end{bmatrix}\\
\end{split}
\end{equation}

\begin{equation}\label{eq:U2_where}
\begin{split}
    &\mbox{where:~}\\    
    &~~~~~~~~~~~~~~~~~~~~a^{2}_{\left|k\right|\left|j\right|} = a_{\left|k\right|}a_{\left|j\right|}\\
    &~~~~~~~~~~~~~~~~~~\varphi_{h,\left|k\right|\left|j\right|} = \tilde{k}\phi_{\left|k\right|} + h \tilde{j}\phi_{\left|j\right|}
\end{split}
\end{equation}

\bibliography{Shih_Modified_Calibration_Function_For_RF_Bunching_Cavity_Bibliography}

\begin{thebibliography}{11}%
\makeatletter
\providecommand \@ifxundefined [1]{%
 \@ifx{#1\undefined}
}%
\providecommand \@ifnum [1]{%
 \ifnum #1\expandafter \@firstoftwo
 \else \expandafter \@secondoftwo
 \fi
}%
\providecommand \@ifx [1]{%
 \ifx #1\expandafter \@firstoftwo
 \else \expandafter \@secondoftwo
 \fi
}%
\providecommand \natexlab [1]{#1}%
\providecommand \enquote  [1]{``#1''}%
\providecommand \bibnamefont  [1]{#1}%
\providecommand \bibfnamefont [1]{#1}%
\providecommand \citenamefont [1]{#1}%
\providecommand \href@noop [0]{\@secondoftwo}%
\providecommand \href [0]{\begingroup \@sanitize@url \@href}%
\providecommand \@href[1]{\@@startlink{#1}\@@href}%
\providecommand \@@href[1]{\endgroup#1\@@endlink}%
\providecommand \@sanitize@url [0]{\catcode `\\12\catcode `\$12\catcode
  `\&12\catcode `\#12\catcode `\^12\catcode `\_12\catcode `\%12\relax}%
\providecommand \@@startlink[1]{}%
\providecommand \@@endlink[0]{}%
\providecommand \url  [0]{\begingroup\@sanitize@url \@url }%
\providecommand \@url [1]{\endgroup\@href {#1}{\urlprefix }}%
\providecommand \urlprefix  [0]{URL }%
\providecommand \Eprint [0]{\href }%
\providecommand \doibase [0]{https://doi.org/}%
\providecommand \selectlanguage [0]{\@gobble}%
\providecommand \bibinfo  [0]{\@secondoftwo}%
\providecommand \bibfield  [0]{\@secondoftwo}%
\providecommand \translation [1]{[#1]}%
\providecommand \BibitemOpen [0]{}%
\providecommand \bibitemStop [0]{}%
\providecommand \bibitemNoStop [0]{.\EOS\space}%
\providecommand \EOS [0]{\spacefactor3000\relax}%
\providecommand \BibitemShut  [1]{\csname bibitem#1\endcsname}%
\let\auto@bib@innerbib\@empty
\bibitem [{\citenamefont {Litvinenko}\ and\ \citenamefont
  {Derbenev}(2009)}]{Litvinenko_CeC}%
  \BibitemOpen
  \bibfield  {author} {\bibinfo {author} {\bibfnamefont {V.~N.}\ \bibnamefont
  {Litvinenko}}\ and\ \bibinfo {author} {\bibfnamefont {Y.~S.}\ \bibnamefont
  {Derbenev}},\ }\bibfield  {title} {\bibinfo {title} {Coherent electron
  cooling},\ }\href@noop {} {\bibfield  {journal} {\bibinfo  {journal}
  {Physical Review Letters}\ }\textbf {\bibinfo {volume} {102}},\ \bibinfo
  {pages} {114801} (\bibinfo {year} {2009})}\BibitemShut {NoStop}%
\bibitem [{\citenamefont {Pinayev}\ \emph {et~al.}(2013)\citenamefont
  {Pinayev}, \citenamefont {Belomestnykh}, \citenamefont {Ben-Zvi},
  \citenamefont {Brown}, \citenamefont {Brutus}, \citenamefont {DeSanto},
  \citenamefont {Elizarov}, \citenamefont {Folz}, \citenamefont {Gassner},
  \citenamefont {Hao} \emph {et~al.}}]{Pinayev_CeC_Status}%
  \BibitemOpen
  \bibfield  {author} {\bibinfo {author} {\bibfnamefont {I.}~\bibnamefont
  {Pinayev}}, \bibinfo {author} {\bibfnamefont {S.}~\bibnamefont
  {Belomestnykh}}, \bibinfo {author} {\bibfnamefont {I.}~\bibnamefont
  {Ben-Zvi}}, \bibinfo {author} {\bibfnamefont {K.}~\bibnamefont {Brown}},
  \bibinfo {author} {\bibfnamefont {J.}~\bibnamefont {Brutus}}, \bibinfo
  {author} {\bibfnamefont {L.}~\bibnamefont {DeSanto}}, \bibinfo {author}
  {\bibfnamefont {A.}~\bibnamefont {Elizarov}}, \bibinfo {author}
  {\bibfnamefont {C.}~\bibnamefont {Folz}}, \bibinfo {author} {\bibfnamefont
  {D.}~\bibnamefont {Gassner}}, \bibinfo {author} {\bibfnamefont
  {Y.}~\bibnamefont {Hao}}, \emph {et~al.},\ }\bibfield  {title} {\bibinfo
  {title} {Present status of coherent electron cooling proof of principle
  experiment},\ }in\ \href@noop {} {\emph {\bibinfo {booktitle} {Proceedings of
  the International Workshop on Beam Cooling and Related Topics, COOL'13,
  WEPPO14}}}\ (\bibinfo {address} {Murren, Switzerland},\ \bibinfo {year}
  {2013})\ pp.\ \bibinfo {pages} {127--129}\BibitemShut {NoStop}%
\bibitem [{\citenamefont {Petrushina}\ \emph {et~al.}(2019)\citenamefont
  {Petrushina}, \citenamefont {Litvinenko}, \citenamefont {Jing},\ and\
  \citenamefont {Ma}}]{petrushina2019measurements}%
  \BibitemOpen
  \bibfield  {author} {\bibinfo {author} {\bibfnamefont {I.}~\bibnamefont
  {Petrushina}}, \bibinfo {author} {\bibfnamefont {V.~N.}\ \bibnamefont
  {Litvinenko}}, \bibinfo {author} {\bibfnamefont {Y.}~\bibnamefont {Jing}},\
  and\ \bibinfo {author} {\bibfnamefont {J.}~\bibnamefont {Ma}},\ }\href@noop
  {} {\emph {\bibinfo {title} {Measurements of the electrical axes of the CeC
  PoP RF cavities}}},\ \bibinfo {type} {Tech. Rep.}\ (\bibinfo  {institution}
  {Brookhaven National Lab.(BNL), Upton, NY (United States)},\ \bibinfo {year}
  {2019})\BibitemShut {NoStop}%
\bibitem [{\citenamefont {Petrushina}\ \emph {et~al.}(2020)\citenamefont
  {Petrushina}, \citenamefont {Litvinenko}, \citenamefont {Jing}, \citenamefont
  {Ma}, \citenamefont {Pinayev}, \citenamefont {Shih}, \citenamefont {Wang},
  \citenamefont {Wu}, \citenamefont {Altinbas}, \citenamefont {Brutus} \emph
  {et~al.}}]{petrushina2020high}%
  \BibitemOpen
  \bibfield  {author} {\bibinfo {author} {\bibfnamefont {I.}~\bibnamefont
  {Petrushina}}, \bibinfo {author} {\bibfnamefont {V.}~\bibnamefont
  {Litvinenko}}, \bibinfo {author} {\bibfnamefont {Y.}~\bibnamefont {Jing}},
  \bibinfo {author} {\bibfnamefont {J.}~\bibnamefont {Ma}}, \bibinfo {author}
  {\bibfnamefont {I.}~\bibnamefont {Pinayev}}, \bibinfo {author} {\bibfnamefont
  {K.}~\bibnamefont {Shih}}, \bibinfo {author} {\bibfnamefont {G.}~\bibnamefont
  {Wang}}, \bibinfo {author} {\bibfnamefont {Y.}~\bibnamefont {Wu}}, \bibinfo
  {author} {\bibfnamefont {Z.}~\bibnamefont {Altinbas}}, \bibinfo {author}
  {\bibfnamefont {J.}~\bibnamefont {Brutus}}, \emph {et~al.},\ }\bibfield
  {title} {\bibinfo {title} {High-brightness continuous-wave electron beams
  from superconducting radio-frequency photoemission gun},\ }\href@noop {}
  {\bibfield  {journal} {\bibinfo  {journal} {Physical Review Letters}\
  }\textbf {\bibinfo {volume} {124}},\ \bibinfo {pages} {244801} (\bibinfo
  {year} {2020})}\BibitemShut {NoStop}%
\bibitem [{\citenamefont {Belomestnykh}\ \emph {et~al.}(2015)\citenamefont
  {Belomestnykh}, \citenamefont {Ben-Zvi}, \citenamefont {Brutus},
  \citenamefont {Litvinenko}, \citenamefont {McIntosh}, \citenamefont {Moss},
  \citenamefont {Narayan}, \citenamefont {Orfin}, \citenamefont {Pinayev},
  \citenamefont {Rao} \emph {et~al.}}]{buncherpaper}%
  \BibitemOpen
  \bibfield  {author} {\bibinfo {author} {\bibfnamefont {S.}~\bibnamefont
  {Belomestnykh}}, \bibinfo {author} {\bibfnamefont {I.}~\bibnamefont
  {Ben-Zvi}}, \bibinfo {author} {\bibfnamefont {J.~C.}\ \bibnamefont {Brutus}},
  \bibinfo {author} {\bibfnamefont {V.}~\bibnamefont {Litvinenko}}, \bibinfo
  {author} {\bibfnamefont {P.}~\bibnamefont {McIntosh}}, \bibinfo {author}
  {\bibfnamefont {A.}~\bibnamefont {Moss}}, \bibinfo {author} {\bibfnamefont
  {G.}~\bibnamefont {Narayan}}, \bibinfo {author} {\bibfnamefont
  {P.}~\bibnamefont {Orfin}}, \bibinfo {author} {\bibfnamefont
  {I.}~\bibnamefont {Pinayev}}, \bibinfo {author} {\bibfnamefont
  {T.}~\bibnamefont {Rao}}, \emph {et~al.},\ }\href@noop {} {\emph {\bibinfo
  {title} {Commissioning of the 112 MHz SRF gun and 500 MHz bunching cavities
  for the CeC PoP linac}}},\ \bibinfo {type} {Tech. Rep.}\ (\bibinfo
  {institution} {Brookhaven National Lab.(BNL), Upton, NY (United States)},\
  \bibinfo {year} {2015})\BibitemShut {NoStop}%
\bibitem [{\citenamefont {Qiang}\ \emph {et~al.}(2006)\citenamefont {Qiang},
  \citenamefont {Lidia}, \citenamefont {Ryne},\ and\ \citenamefont
  {Limborg-Deprey}}]{qiang2006three}%
  \BibitemOpen
  \bibfield  {author} {\bibinfo {author} {\bibfnamefont {J.}~\bibnamefont
  {Qiang}}, \bibinfo {author} {\bibfnamefont {S.}~\bibnamefont {Lidia}},
  \bibinfo {author} {\bibfnamefont {R.~D.}\ \bibnamefont {Ryne}},\ and\
  \bibinfo {author} {\bibfnamefont {C.}~\bibnamefont {Limborg-Deprey}},\
  }\bibfield  {title} {\bibinfo {title} {Three-dimensional quasistatic model
  for high brightness beam dynamics simulation},\ }\href@noop {} {\bibfield
  {journal} {\bibinfo  {journal} {Physical Review Special Topics-Accelerators
  and Beams}\ }\textbf {\bibinfo {volume} {9}},\ \bibinfo {pages} {044204}
  (\bibinfo {year} {2006})}\BibitemShut {NoStop}%
\bibitem [{\citenamefont {Delayen}(1987)}]{delayen1987longitudinal}%
  \BibitemOpen
  \bibfield  {author} {\bibinfo {author} {\bibfnamefont {J.}~\bibnamefont
  {Delayen}},\ }\bibfield  {title} {\bibinfo {title} {Longitudinal transit time
  factors of short independently phased accelerating structures for low
  velocity ions},\ }\href@noop {} {\bibfield  {journal} {\bibinfo  {journal}
  {Nuclear Instruments and Methods in Physics Research Section A: Accelerators,
  Spectrometers, Detectors and Associated Equipment}\ }\textbf {\bibinfo
  {volume} {258}},\ \bibinfo {pages} {15} (\bibinfo {year} {1987})}\BibitemShut
  {NoStop}%
\bibitem [{\citenamefont {Carne}\ \emph {et~al.}(1970)\citenamefont {Carne},
  \citenamefont {Lapostolle}, \citenamefont {Schnizer},\ and\ \citenamefont
  {Prome}}]{carne1970numerical}%
  \BibitemOpen
  \bibfield  {author} {\bibinfo {author} {\bibfnamefont {A.}~\bibnamefont
  {Carne}}, \bibinfo {author} {\bibfnamefont {P.}~\bibnamefont {Lapostolle}},
  \bibinfo {author} {\bibfnamefont {B.}~\bibnamefont {Schnizer}},\ and\
  \bibinfo {author} {\bibfnamefont {M.}~\bibnamefont {Prome}},\ }\href@noop {}
  {\emph {\bibinfo {title} {NUMERICAL METHODS: ACCELERATION BY A GAP.}}},\
  \bibinfo {type} {Tech. Rep.}\ (\bibinfo  {institution} {Rutherford High
  Energy Lab., Chilton, Eng.},\ \bibinfo {year} {1970})\BibitemShut {NoStop}%
\bibitem [{\citenamefont {Jing}\ \emph {et~al.}(2021)\citenamefont {Jing},
  \citenamefont {Litvinenko}, \citenamefont {Pinayev}, \citenamefont {Wu},
  \citenamefont {Petrushina},\ and\ \citenamefont {Shih}}]{jing2021beam}%
  \BibitemOpen
  \bibfield  {author} {\bibinfo {author} {\bibfnamefont {Y.}~\bibnamefont
  {Jing}}, \bibinfo {author} {\bibfnamefont {V.~N.}\ \bibnamefont
  {Litvinenko}}, \bibinfo {author} {\bibfnamefont {I.}~\bibnamefont {Pinayev}},
  \bibinfo {author} {\bibfnamefont {Y.}~\bibnamefont {Wu}}, \bibinfo {author}
  {\bibfnamefont {I.}~\bibnamefont {Petrushina}},\ and\ \bibinfo {author}
  {\bibfnamefont {K.}~\bibnamefont {Shih}},\ }\bibfield  {title} {\bibinfo
  {title} {Beam dynamics in coherent electron cooling accelerator},\
  }\href@noop {} {\bibfield  {journal} {\bibinfo  {journal} {Nuclear
  Instruments and Methods in Physics Research Section A: Accelerators,
  Spectrometers, Detectors and Associated Equipment}\ } (\bibinfo {year}
  {2021})}\BibitemShut {NoStop}%
\bibitem [{\citenamefont {Brutus}\ \emph {et~al.}(2014)\citenamefont {Brutus},
  \citenamefont {Hulsart}, \citenamefont {Litvinenko}, \citenamefont
  {Michnoff}, \citenamefont {Miller}, \citenamefont {Minty}, \citenamefont
  {Pinayev},\ and\ \citenamefont {Wilinski}}]{brutus2014coherent}%
  \BibitemOpen
  \bibfield  {author} {\bibinfo {author} {\bibfnamefont {J.}~\bibnamefont
  {Brutus}}, \bibinfo {author} {\bibfnamefont {R.}~\bibnamefont {Hulsart}},
  \bibinfo {author} {\bibfnamefont {V.}~\bibnamefont {Litvinenko}}, \bibinfo
  {author} {\bibfnamefont {R.}~\bibnamefont {Michnoff}}, \bibinfo {author}
  {\bibfnamefont {T.}~\bibnamefont {Miller}}, \bibinfo {author} {\bibfnamefont
  {M.}~\bibnamefont {Minty}}, \bibinfo {author} {\bibfnamefont
  {I.}~\bibnamefont {Pinayev}},\ and\ \bibinfo {author} {\bibfnamefont
  {M.}~\bibnamefont {Wilinski}},\ }\href@noop {} {\emph {\bibinfo {title}
  {Coherent electron cooling proof of principle phase 1 instrumentation
  status}}},\ \bibinfo {type} {Tech. Rep.}\ (\bibinfo  {institution}
  {Brookhaven National Lab.(BNL), Upton, NY (United States). Relativistic
  Heavy~…},\ \bibinfo {year} {2014})\BibitemShut {NoStop}%
\bibitem [{\citenamefont {Pinayev}\ \emph {et~al.}(2021)\citenamefont
  {Pinayev}, \citenamefont {Jing}, \citenamefont {Kayran}, \citenamefont
  {Litvinenko}, \citenamefont {Ma}, \citenamefont {Mihara}, \citenamefont
  {Petrushina}, \citenamefont {Shih}, \citenamefont {Wang},\ and\ \citenamefont
  {Wu}}]{igorEmeasurement}%
  \BibitemOpen
  \bibfield  {author} {\bibinfo {author} {\bibfnamefont {I.}~\bibnamefont
  {Pinayev}}, \bibinfo {author} {\bibfnamefont {Y.}~\bibnamefont {Jing}},
  \bibinfo {author} {\bibfnamefont {D.}~\bibnamefont {Kayran}}, \bibinfo
  {author} {\bibfnamefont {V.~N.}\ \bibnamefont {Litvinenko}}, \bibinfo
  {author} {\bibfnamefont {J.}~\bibnamefont {Ma}}, \bibinfo {author}
  {\bibfnamefont {K.}~\bibnamefont {Mihara}}, \bibinfo {author} {\bibfnamefont
  {I.}~\bibnamefont {Petrushina}}, \bibinfo {author} {\bibfnamefont
  {K.}~\bibnamefont {Shih}}, \bibinfo {author} {\bibfnamefont {G.}~\bibnamefont
  {Wang}},\ and\ \bibinfo {author} {\bibfnamefont {Y.~H.}\ \bibnamefont {Wu}},\
  }\bibfield  {title} {\bibinfo {title} {Using solenoid as multipurpose tool
  for measuring beam parameters},\ }\href@noop {} {\bibfield  {journal}
  {\bibinfo  {journal} {Review of Scientific Instruments}\ }\textbf {\bibinfo
  {volume} {92}},\ \bibinfo {pages} {013301} (\bibinfo {year}
  {2021})}\BibitemShut {NoStop}%
\end{thebibliography}%

\end{document}